\documentclass[10pt,letterpaper]{article}
\usepackage{opex3}

\begin{document}


\title{Real-time control of the periodicity of a standing wave: an optical accordion}

\author{T. C. Li, H. Kelkar, D. Medellin, and M. G. Raizen}

\address{Center for Nonlinear Dynamics and Department of Physics,
         \\The University of Texas at Austin, Austin, Texas 78712, USA}

\email{tcli@physics.utexas.edu} 



\begin{abstract}
We report an experimental method to create optical lattices with
real-time control of their periodicity. We demonstrate a continuous
change of the lattice periodicity from 0.96 $\mu$m to 11.2 $\mu$m in
one second, while the center fringe only moves less than 2.7 $\mu$m
during the whole process. This provides a powerful tool for
controlling ultracold atoms in optical lattices, where small spacing
is essential for quantum tunneling, and large spacing enables
single-site manipulation and spatially resolved detection.
\end{abstract}



\section{Introduction}

Ultracold atoms in optical lattices are a model system for studying
quantum many-body effects in a highly controllable way
\cite{morsch06}. They are also promising candidates for quantum
information processing \cite{brennen99, mandel03}. Recently, optical
lattices with small periodicity have been used to investigate
superfluidity \cite{cataliotti01},  quantum transport
\cite{henderson06}, the superfluid-Mott insulator transition
 \cite{greiner02,gerbier06}, and the Tonks-Girardeau
gas \cite{paredes04,kinoshita04}. Small periodicity is essential for
these experiments as high  tunneling rates between neighboring sites
are required. However,  small periodicity also makes it very
difficult to manipulate and detect  single sites of the optical
lattice. For example, the number statistics of a single site of the
Mott insulator phase has never been measured directly.  On the other
hand, single-site manipulation and detection has been demonstrated
in optical lattices with large periodicity \cite{scheunemann00,
nelson07}. We can bridge this gap if we are able to change the
lattice periodicity while keeping atoms trapped  in the lattice.

Several groups have  created optical lattices with tunable spacing
\cite{peil03,fallani05,venkatakrishnan03,tan05,xie07}, however, a
configuration that is stable enough to allow continuous variation of
the lattice periodicity while keeping  atoms trapped  has yet to be
developed. As the optical lattices are normally created by
interfering two beams, which is sensitive to the relative phase
between beams, it is very difficult to keep the lattice stable while
changing its spacing. In this paper, we report the creation of an
optical lattice formed by two parallel
 beams brought together at the focal plane of a lens. We use a novel
method to change the distance between the two beams that keeps the
 optical lattice stable. The center fringe shifts less than
2.7 $\mu$m while the lattice spacing is changed from 0.96 $\mu$m to
11.2 $\mu$m in one second. This stability   allows one to  change
the lattice spacing in real time while keeping atoms trapped. This
is also  useful in many other applications of optical lattices, such
as sorting microscopic particles \cite{macdonald03}.

\section{Theory and Experiment}

\begin{figure}[htb]
\centering\includegraphics[width=6cm]{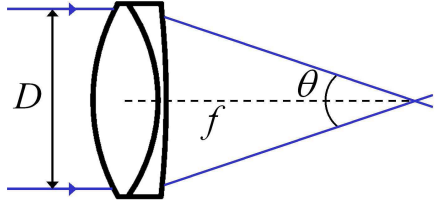} \caption{\label{fig1}
Two parallel
  beams separated by a distance $D$, produce an optical lattice at the
focal plane of the lens. We use an achromatic doublet lens to
minimize the aberration of the off-axis beams. The highly curved
surface faces the incident parallel beams.}
\end{figure}

As shown in Fig. \ref{fig1}, two parallel beams separated by a
distance $D$ are brought together by a lens to produce an optical
lattice  at the focal plane of the lens. If the beams at the focus
are plane waves, the lattice spacing will be $d=\lambda/(2
\sin(\theta/2))$, where $\lambda$ denotes the laser wavelength, and
$\theta$ is the angle between the two beams at the focus. For a thin
lens, if the beams are symmetric with respect to the axis of the
lens, the angle will be approximately $\theta \approx 2
\tan^{-1}(D/2f)$. Thus
\begin{equation}
\label{eq1} d \approx \lambda \frac{(D^2/4+f^2)^{1/2}}{D},
\end{equation}
where  $f$ is the focal length of the lens. We can tune the lattice
spacing by changing $D$. Eq. (\ref{eq1}) is commonly used for
calculating the lattice spacing \cite{venkatakrishnan03}, however,
it is not very accurate when  $\theta$ is large.

The interference pattern of the two beams and its lattice spacing
can be calculated more precisely using Fourier optics. For our
setup, we use an achromatic doublet lens, which has much smaller
aberrations than a singlet lens, and can be used to achieve a
diffraction limited laser spot. It  performs a Fourier
transformation when an object sits at the front focal plane and the
image sits at the back focal plane. Assuming that the field of an
incident laser at the front focal plane of a lens is $U(x_1,\, y_1,
\,-f)$, then the field at the back focal plane of the lens $U_0(x,
\, y, \, f)$ is the Fourier transform of $U(x_1,\, y_1, \,-f)$.
 For a Gaussian beam, the resulting intensity distribution
$I_0(x, \, y, \, f)=U_0^*(x, \, y, \, f) \, U_0(x, \, y, \, f)$ is
still Gaussian and can be calculated easily. If the incident beam is
shifted horizontally by $+D/2$ or $-D/2$, the resulted field at the
back focal plane will become
\begin{eqnarray}
\label{eq3}
&&U_{+D/2}(x, \, y, \, f)= \exp({-j\, \frac{\pi D}{\lambda f} x}) \, U_0(x, \, y, \, f),\\
\label{eq4} {\textrm{or}}&&U_{-D/2}(x, \, y, \, f)= \exp({+j\,
\frac{\pi D}{\lambda f} x}) \, U_0(x, \, y, \, f).
\end{eqnarray}

Thus the interference pattern created by bringing two identical
parallel beams together by a lens (see Fig. \ref{fig1}) is
\begin{equation}
\label{eq5} I(x, \, y, \, f)=2 \, (\cos \frac{2 \pi D}{\lambda f}x
+1) \, I_0(x, \, y, \, f),
\end{equation}
and the period of the interference pattern is
\begin{equation}
\label{eq6} d= \lambda f/D\, .
\end{equation}

As shown in the derivation, Eq. (\ref{eq5}) is independent of the
curvature of the laser beams. Both beams can be divergent and have
very large waists at the focal plane. Thus we can use this method to
produce optical lattices with arbitrary size.  Moreover, Eq.
(\ref{eq6}) is much simpler and more accurate than Eq. (\ref{eq1})
when  $\theta$ is large.  From Eq. (\ref{eq6}), we  get the  angle
between the two beams at the focus to be $\theta = 2
\sin^{-1}(D/2f)$.

\begin{figure}[htb]
\centering\includegraphics[width=12cm]{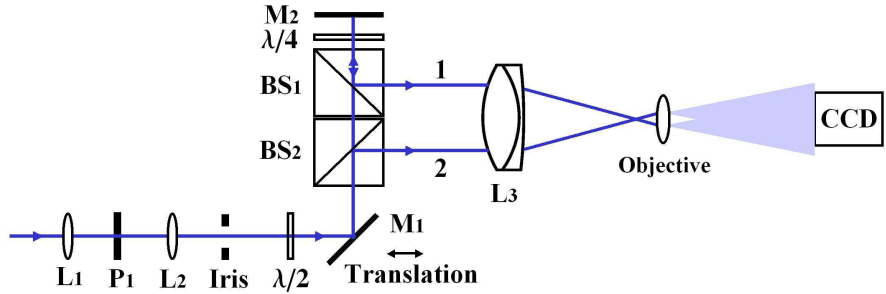} \caption{\label{fig2}
Experimental set-up for creating and imaging the optical lattices
with real-time control of periodicity. L{\scriptsize 1},
L{\scriptsize 2} are singlet lenses; P{\scriptsize 1} is a pinhole;
 M{\scriptsize 1} and M{\scriptsize 2} are  mirrors; BS{\scriptsize 1} and BS{\scriptsize 2}
  are polarizing cube beam splitters;
 and L{\scriptsize 3} is an achromatic doublet lens. }
\end{figure}

Our experimental setup for creating and imaging the optical lattices
is  shown  in Fig. \ref{fig2}.
 A Gaussian beam with the
desired waist is obtained by passing a laser beam ($\lambda=532$ nm)
through two lenses (L{\scriptsize 1}, L{\scriptsize 2}), one pinhole
(P{\scriptsize 1}) and one iris.  A $\lambda /2$ waveplate is used
to change the orientation of the polarization of the beam, which
controls the relative power of the two parallel beams. The
s-polarized component of the beam is reflected by BS{\scriptsize 2}
and the p-polarized component passes through both beamsplitters.
After passing through the $\lambda/4$ waveplate twice, the
p-polarized component becomes s-polarized and is reflected by
BS{\scriptsize 1}. These two parallel beams are brought together by
an achromatic doublet lens L{\scriptsize 3} to create the optical
lattice. Then the optical lattice is magnified by an objective and
imaged by a CCD camera. We  tried two lenses with different focal
lengths (30.0 and 80.0 mm) for L{\scriptsize 3}.  The optical
lattice is formed at the focal plane of
 L{\scriptsize 3}. Its periodicity can be changed in real time by moving M{\scriptsize
 1} with a stepper motor,
and its fringe contrast can be close to 100\%
 by tuning the $\lambda /2$ waveplate.

\begin{figure}[bht]
\centering\includegraphics[width=6.5cm]{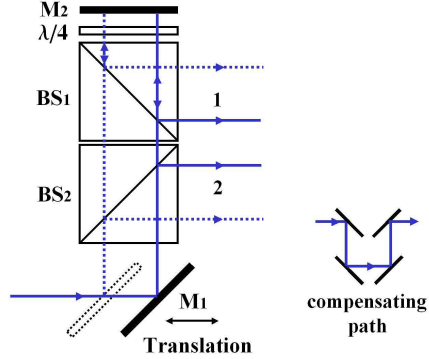} \caption{\label{fig3}
The distance between the two parallel beams can be changed by moving
M1. }
\end{figure}

As shown in Fig. \ref{fig3}, the distance between the two parallel
beams can be changed by moving mirror M{\scriptsize 1} horizontally
or vertically. The difference between the optical path lengths of
the two beams does not change when M{\scriptsize 1} is moving. Thus
in principle the center fringe of the optical lattice will not move
when  M{\scriptsize 1} is moving in any direction. A compensating
path, as shown in the right part of Fig. \ref{fig3}, can be used to
make the difference of the optical path lengths to be zero, if
necessary, for lasers with small coherence length or divergent
beams. The optical lattice is also not sensitive to the alignment of
the beamsplitters. The objective lens is aligned by making sure that
the image of  beam 2 does not move when M{\scriptsize 1} does,
ensuring that the focal plane of L{\scriptsize 3} is imaged at the
CCD camera. Then we adjust M{\scriptsize 2} to let the image of beam
1 superpose with beam 2, which automatically makes two beams
parallel.

\section{Results and discussion}

\begin{figure}[htb]
\centering\includegraphics[width=6cm]{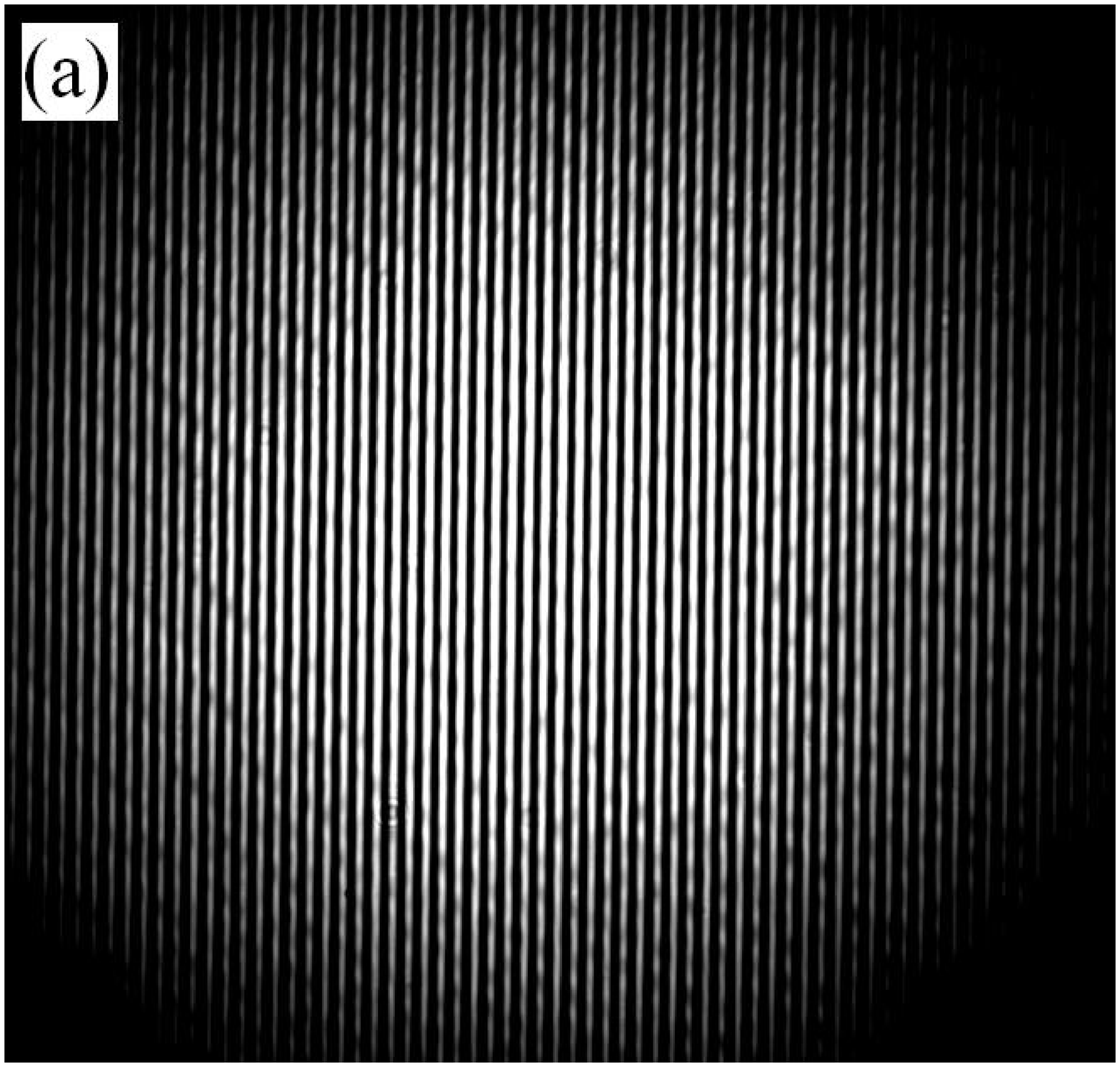}
\centering\includegraphics[width=6.6cm]{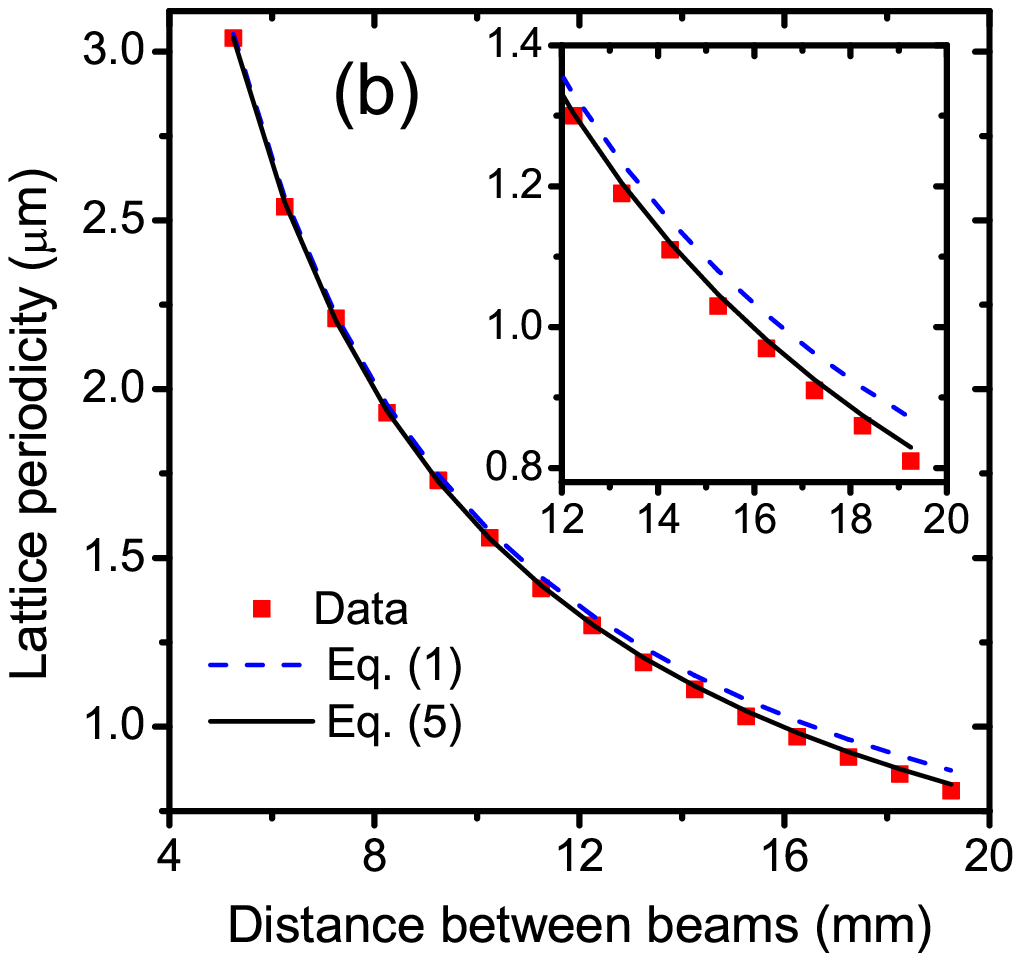} \caption{\label{fig4}
(a): An optical lattice with spacing of 0.81 $ \mu$m, recorded by a
CCD camera; (b): the lattice periodicity at the focus of a $f=30$ mm
lens as a function of the distance between the two parallel beams.
The statistical error of the data is smaller than the size of the
squares.}
\end{figure}

An image of the  interference pattern created at the focus of the
lens is shown in Fig. \ref{fig4}(a). The
 periodicity of this optical lattice is 0.81 $ \mu$m. It was created from  two parallel beams
separated by 19.25 mm, which interfered at the focal plane of a
$f=30.0$ mm lens. The waists of the two beams at the focus are 36
$\mu$m and 40 $\mu$m, which were measured by a scanning knife edge.
Although the beams were close to the edge of the lens and the angle
was large ($\theta \approx 38^{\circ}$), the interference fringes
were very straight even at the edge of the beams. This agrees with
Eq. (\ref{eq6}).

 Figure
\ref{fig4}(b) shows the lattice spacing as a function of the
distance between the two parallel beams. The lattice spacings are
determined by Fourier transforming the images. The imaging system
can be calibrated by the waists of the beams, which gives
$0.0853\pm0.0053 \mu$m/pixel, where the uncertainty comes from the
measurements of beam waists. It can also be calibrated by fitting
the data with Eq. (\ref{eq6}), which gives $0.0855\pm0.0001
\mu$m/pixel. We use $0.0853 \mu$m/pixel for the data  shown in the
figure. Equation (\ref{eq6}) agrees with all data points, and
clearly fits the data better than Eq. (\ref{eq1}) when $D$ is large.

\begin{figure}[htb]
\centering\includegraphics[width=8cm]{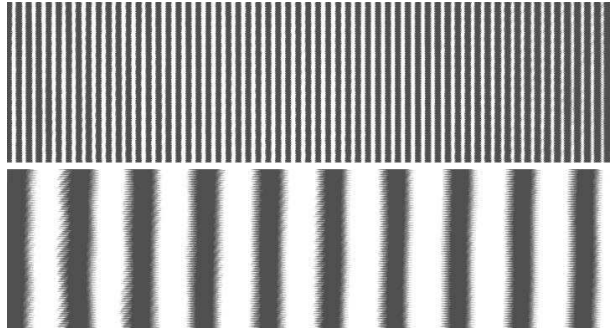} \caption{\label{fig5}
 Optical lattices with  spacing of 0.98 $
\mu$m and 6.20 $ \mu$m. }
\end{figure}

Figure \ref{fig5} shows the optical lattices with spacing of 0.98 $
\mu$m and 6.20 $ \mu$m. They are formed at the focal plane of a
$f=80.0$ mm lens. The change of spacing is achieved by moving
M{\scriptsize 1}.

\begin{figure}[htb]
\centering\includegraphics[width=6cm]{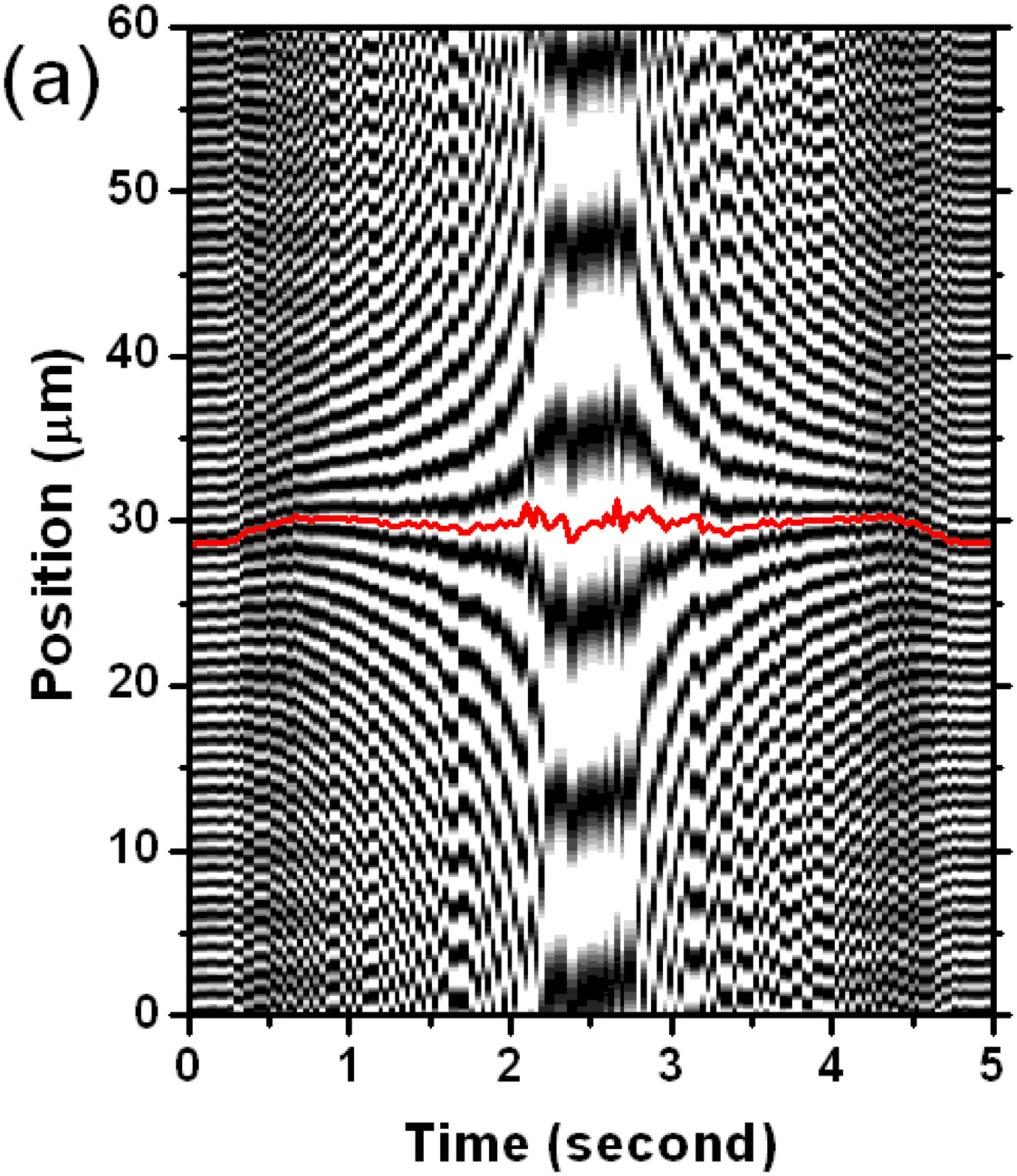}
\centering\includegraphics[width=6cm]{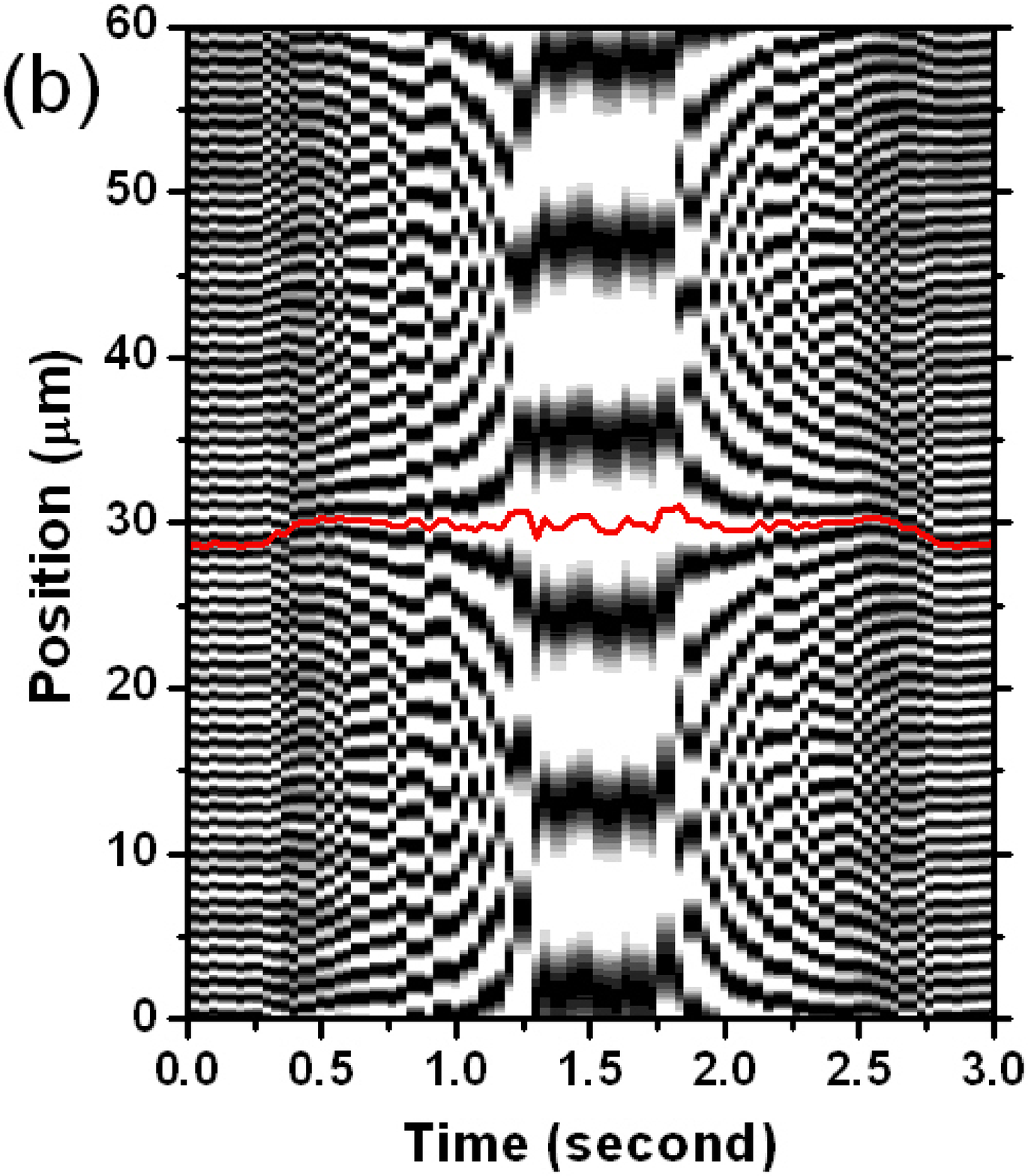}
 \caption{\label{fig6} Continuous change of the lattice periodicity from 0.96 $\mu$m to 11.2 $\mu$m,
 and back to 0.96 $\mu$m by moving M{\scriptsize 1} at different speed: 10 mm/s (a) and 20 mm/s (b).
  The center fringe (marked by a solid line)
 moved less than 2.7 $\mu$m during the whole process.  }
\end{figure}

The continuous change of the lattice periodicity from 0.96 $\mu$m to
11.2 $\mu$m, and back to 0.96 $\mu$m  in real time  is presented in
Fig. \ref{fig6}.  The optical lattice is formed at the focal plane
of a $f=80.0$ mm lens. Mirror M{\scriptsize 1} is moved horizontally
by 20.0 mm to change $D$ from 43.81 mm to 3.79 mm. This figure is
constructed from real-time images (30 frames/second) similar to Fig.
\ref{fig5}. It shows that our optical lattice is very stable. The
center fringe (marked by a solid line)
  moved less than 2.7 $\mu$m during the whole process, which is
 only quarter of the final lattice spacing.  There is no apparent difference in the vibration of the lattice
 whether M{\scriptsize 1} is moving or not. This means that the vibration due to translating M{\scriptsize 1}
  does not transfer  to
 the optical lattice. In Fig. \ref{fig6}(b), we change the lattice spacing
from 0.96 $\mu$m to 11.2 $\mu$m in one second, wait for   half a
second, and change the spacing back to 0.96 $\mu$m in another one
second. The life time of ultracold atoms in the optical lattices can
be longer than 10 seconds \cite{nelson07}, so this method enables
one to change the lattice spacing in real time while keeping atoms
trapped.

 We
also  changed  $D$ by moving BS{\scriptsize 1} and BS{\scriptsize 2}
together horizontally, which was similar to Ref. \cite{tan05}. In
this case, the center fringe shifted much more and  was extremely
sensitive to the vibrations. If the motion is not exactly straight
and parallel to the beams 1 and 2 (see Fig. \ref{fig2}), but
deviates (or vibrates) by only 2.13 $\mu$m in the perpendicular
direction, the difference between the optical path lengths of the
two beams will change by 4.26 $\mu$m. Thus the center fringe of the
optical lattices will shift (or vibrate) by 8 fringes, which is 80
$\mu$m when the lattice spacing is 10 $\mu$m.

\section{Conclusion}
In this paper we have presented an experimental method to create
optical lattices with real time control of periodicity. The center
fringe of the optical lattice shifts less than 2.7 $\mu$m while the
lattice spacing is changed by  one order of magnitude.  Such
accordion lattices can work as magnifiers or compressors for many
applications.

\section*{Acknowledgments}
The authors would like to acknowledge support from the Sid W.
Richardson Foundation, the National Science Foundation, and the R.
A. Welch Foundation.

\end{document}